# The Monster Sporadic Group and a Theory Underlying Superstring Models

George Chapline

Physics and Space Technology Division
Lawrence Livermore National Laboratory
Livermore, Ca 94550
chapline1@llnl.gov

## Abstract

The pattern of duality symmetries acting on the states of compactified superstring models reinforces an earlier suggestion that the Monster sporadic group is a hidden symmetry for superstring models. This in turn points to a supersymmetric theory of self-dual and anti-self-dual K3 manifolds joined by Dirac strings and evolving in a 13 dimensional spacetime as the fundamental theory. In addition to the usual graviton and dilaton this theory contains matter-like degrees of freedom resembling the massless states of the heterotic string, thus providing a completely geometric interpretation for ordinary matter.

# 1.Introduction

While superstring models possess many mathematically interesting properties, they are not yet satisfactory as fundamental theories of nature for several reasons: a) superstring models are only defined perturbatively, and it is not clear whether non-perturbative formulations of the theory exist, b) superstring models seem to require for their formulation a classical background spacetime whose very existence is inconsistent with a quantum theory of gravity in four dimensions, c) there are numerous varieties of superstring models, and no principle has yet been discovered which would compel one to single out a single variety as the one which corresponds to the world we live in, and d) heterotic superstring models do not provide a geometrical interpretation for ordinary matter degrees of freedom. Of course it is possible that all the superstring models represent different manifestations of a single underlying theory, and in fact much evidence has recently accumulated to support such a suggestion (see for example [1] [2] [3] ). In this paper we wish to point out that a refinement of a finite and exactly solvable quantum model for four dimensional spacetime previously introduced by the author [4] seems to be a good candidate for the underlying theory.

The connecting link between superstring models and the quantum model for four dimensional spacetime is the largest sporadic finite group $F_1$, often referred to as the "Monster" group [5]. Because of the structure of the anomaly cancellation mechanism in superstring models the author suggested some time ago [6] that there is an underlying theory for heterotic and Type II superstring models with the Monster sporadic group as its symmetry. Remarkably the evidence which has recently accumulated in favor of an single theory underlying all superstring models also seems to support this suggestion. For example, the discovery of a string-string duality symmetry between Type II and heterotic superstring models in six dimensions [7] points to an underlying symmetry that unifies these two types of superstrings. As pointed out in reference 6 the structure of the algebra which forms smallest nontrivial representation of the Monster group suggests that the Monster group transforms Type II superstring states into heterotic string states and visa versa.

Perhaps even more impressive evidence for the presence of the Monster comes from the fact that the duality symmetries of the six dimensional Type II and heterotic string models that are equivalent under string-string duality are strikingly similar to elementary group structures which play a crucial role in the canonical constructions of the Monster group [5] [8] [9]. In particular, the T-duality group SO(4,20) for these string models implies that there are 24 different U(1) groups that can act on the BPS spectrum. What is important for us to notice is that there is a correspondence between these U(1) subgroups of SO(4,20) and the 24 roots of SO(8). Because of the celebrated triality of the $D_4$ root diagram, pairs of the 24 roots of SO(8) can be summed together in groups of four to form the generators of six Cartan subalgebras. The SO(8) Lie algebra can be decomposed into a sum of the conventional Cartan subalgebra and these six other Cartan subalgebras. Any two of the seven Cartan subalgebras generate a copy of $(SL_2)^4$, and therefore one can think of the decomposition of the $D_4$ root diagram into three root systems of type $(SL_2)^4$ as arising from three possible "flavors" for the $SL_2$ generators associated with each Cartan subalgebra. This $SL_2$ triality symmetry can be combined with the $S_3$ triality of the $D_4$ root diagram to yield an $L_3(2)$ symmetry which interchanges the seven Cartan subalgebras of the SO(8) Lie algebra. Remarkably the action of this $L_3(2)$ symmetry on the Cartan subalgebras of the SO(8) Lie algebra exactly mirrors the action of S, T, U triality [10] and string-string duality on the SO(4,20) charge lattice of the dual six dimensional heterotic and Type II superstring models. Now it happens that the decomposition of the SO(8) algebra into seven Cartan subalgebras induces a decomposition of the Griess algebra into seven subalgebras that are permuted by the same type of $L_3(2)$ symmetry [9]. Taken together with the identification of these subalgebras with states of the heterotic and Type II superstrings this is a strong hint that the Monster group plays some role in determining the duality symmetries of compactified superstring models.

The duality symmetries of the six dimensional heterotic and Type II superstring models that are equivalent under string-string duality seem to be associated with the quaternionic construction of the Leech lattice. In a similar way one can argue that the duality symmetries of the eight dimensional heterotic and Type II superstring models that are equivalent under string-string duality seem to be associated with the construction of the Leech lattice from eight SU(4) root lattices. Curiously though there does not seem to be any obvious relation between 10-dimensional superstring theories and a nice construction of the Leech lattice. Instead on the basis of the apparent relationship between the dual superstring models in six and eight dimensions and nice constructions of the Leech lattice one would guess that the connection between superstrings and the Leech lattice would find its most natural setting in 12-dimensions. The standard constructions of the Monster group [5] [8] are closely related to the construction the Leech lattice from 24 copies of the $A_1$ root lattice by the addition of vectors whose coordinates are identical with the binary words of the length 24 Golay code. However an exact duplicate of this construction exists in 12 dimensions in which the complex Leech lattice is constructed from twelve copies of the $A_2$ root lattice by the addition of vectors identified with words of the ternary Golay code. Since the duality symmetries of superstring models in six and eight (and incidently also four) dimensions point to 12-dimensions we are led to suspect that there is some 12-dimensional formulation of string theory for which the connection between superstrings, the complex Leech lattice, and the Monster group becomes evident. The remainder of this paper will be devoted to an attempt to elucidate this 12-dimensional theory.

Athough a general consensus now exists that there should be a theory underlying superstring models, the nature of the degrees of freedom of this theory has remained obscure. Our suggestion in this connection is motivated by Griess' construction of the the Monster sporadic group as the automorphism group of a particular commutative, non associative algebra [5]. In particular Griess' construction suggests that the degrees of freedom of an underlying theory should be represented by commuting but non-associative variables. The elementary quantum mechanical model that this brings to mind is a two dimensional system of nonrelativistic fermions moving in the magnetic potential of a solenoid. Thus even though the quantum Hall effect would not be directly relevant, models of anyonic superfluids [11] might well be good models for the underlying theory. In fact starting from a 3-dimensional generalization of the anyonic superfluid [12] the author has introduced [4] an exactly solvable quantum model for four dimensional spacetime that seems to utilize the type of variables called for in a theory underlying superstring models. The author had previously argued [13] from various general points of view that one way to think about the nature of the degrees of freedom of the theory underlying superstring models is to regard the underlying theory as a quantum theory of four dimensional gravity, and it is now our contention that a refinement of the quantum model for spacetime introduced in ref.4 is the underlying theory.

## 2. A Model for the Quantum Gravity Vacuum

In the quantum model of spacetime introduced in ref.4 the underlying theory of massless fermions interacting via a gauge field has soliton solutions that can be identified as the ground state wavefunction for a system of interacting self-dual and anti-self-dual gravitons. In particular, the following pairing wavefunction seems to be a good candidate for the ground state of four dimensional quantum gravity:

$$\Psi_0 = \prod_{i<j}(\bar{z}_i - \bar{z}_j)(\bar{z}_{[i]} - \bar{z}_{[j]}) \cdot \left[\frac{R_{ij} - U_{ij}}{R_{ij} + U_{ij}}\right]^{1/2} \left[\frac{R_{[i][j]} - U_{[i][j]}}{R_{[i][j]} + U_{[i][j]}}\right]^{1/2} \prod_{k,l}\left[\frac{R_{k[l]} - U_{k[l]}}{R_{k[l]} + U_{k[l]}}\right]^{1/2} \quad (1)$$

where $(z_i, U_i)$ and $(\bar{z}_i, U_i)$ are the positions of the self-dual and anti-self-dual gravitons respectively, $R_{ij}^2 = U_{ij}^2 + (z_i - z_j)(\bar{z}_i - \bar{z}_j)$, and $\sum_{\chi=0}$ is a sum over all sets of self-dual and anti-self-dual gravitons such that the overall Euler characteristic is zero. As shown in ref.4 the wavefunction (1) satisfies a Wheeler-DeWitt like equation $H\Psi=0$, where the Hamiltonian H has an N = 2 supersymmetric form H = $Q_j^+ Q_j$, where $Q_j = D_{x_j} - iD_{y_j}$. In terms of the positions $X_j = (z_i, U_i)$ and momenta $\Pi_j$ of the gravitons the hamitonian H takes the form :

$$H = \sum_j \left(\Pi_j \cdot \Pi_j - B_j\right) \quad (2)$$

where $\Pi_j = -i\nabla_j + A_j$ and $B_j = \sum_k \partial_{U_j} \frac{1}{|r_j - r_k|}$. This hamitonian is the same as that for a collection of nonrelativistic spin-aligned particles with g=2 interacting via a magnetic monopole potential. The Wheeler-DeWitt equation $H\Psi=0$ will have interesting solutions if there are non-trivial solutions to $Q_j \psi = 0$. The existence of non-trivial solutions to $Q_j \psi = 0$ is an example of the Atiyah- Singer index theorem, and generalizes the well known zero modes of charged fermions moving in two dimensions in an arbitrary magnetic field to a model of massless anyons interacting via an SU(∞) gauge field [12]. Our interpretation of the SU(∞) anyons in this theory is that they represent "gravitons". These gravitons resemble the chiral fields of a 2-dimensional conformal field theory; in particular, solutions to eq. (2) representing self-dual or anti-self-dual gravitons obey the same type of fusion rules as the primary chiral fields of an N = 2 superconformal field theory. In fact, this correspondence between gravitons and N = 2 chiral fields is just another way of viewing the magical classification of N = 2 superconformal field theories using singularity theory [14].

The ground state corresponding to (1) possesses the same type of off-diagonal long range order as an anyonic superfluid. The trick to exhibiting the long range order in an anyonic superfluid, originally introduced by Girvin and MacDonald [15], is to make a singular gauge transformation that transforms the anyonic wavefunction into a wavefunction for interacting bosons. In the case of the wavefunction (1) the order parameter can be thought of as the energy of two impurity monopole-antimonopole pairs separated by a long distance. The energy of test monopoles with unit magnetic charges located at $r_1,...,r_n$ is given by

$$\Phi(r_1,...,r_n) = \sum_{l=1}^{n} \sum_{k} \frac{m_k}{|r_l - r_k|}, \qquad (3)$$

where the $m_j = \pm 1$ are the condensate monopole charges. The existence of long range order is equivalent to the fact that $\Phi(r_1,r_{[1]};r_2,r_{[2]})$ for $r_1 \approx r_{[1]}$ and $r_2 \approx r_{[2]}$ approaches a constant as $|r_1 - r_2|$ becomes large. It might be noted that the introduction of a condensate wavefunction involving monopole pairs as the ground state for four dimensional quantum gravity brings quantum gravity more into alignment with nonabelian gauge theories, where it is known that at least in some circumstances the ground state is a monopole condensate [16]. Indeed, quite apart from the question as to whether the particular pairing wavefunction introduced in ref. 4 is a good approximation for classical spacetime, exhibition of this wavefunction gives flight to the idea that ordinary four dimensional classical spacetime corresponds to a condensate of solitons.

It can be shown [12] that the self-dual and anti-self- dual pieces of the wavefunction (1) are closely related to the coefficients of the Gibbons- Hawking metric [17] for an asymptotically locally flat self-dual or anti-self-dual Einstein space. In particular in x,y,U coordinates the Gibbons- Hawking metric has the form:

$$ds^2 = \Phi_+ dx \cdot dx + \frac{1}{\Phi_+}(d\tau + \vec{\omega} \cdot dx)^2, \qquad (4)$$

where $\text{grad}\,\Phi_+ = \text{curl}\,\vec{\omega}$ and $\Phi_+$ is the magnetostatic potential due to magnetic monopoles with positive charges located at positions $X_i$. In order to avoid singularities in the vector potential $\vec{\omega}$ the monopole singularities in $\Phi_+$ must be accompanied by Dirac strings. The effect of these Dirac strings running to infinity is to change the global topology of spacetime at infinity so that instead of being globally flat it has the topology $S/Z_n \times R$, where $Z_n$ is a finite cyclic subgroup of SO(3) [17]. On the other hand when self-dual and anti-self-dual gravitons are paired as in the wavefunction (1), the Dirac strings can run between the paired gravitons and so asymptotically the global topology will again be trivial. Locally though the topology will be nontrivial and will be determined by the small separation of the paired self-dual and anti-self-dual gravitons. It is implicit in the form of the Gibbons-Hawking metric (and explicitly evident in the case of the Eguchi-Hanson metric) that each self-dual or anti-self-dual graviton in eq.(1) has an "orientation" described by the vector $X_{i+1}$-$X_i$. Since the order parameter corresponding to eq. (3) involves the pairing of the self-dual and anti-self-dual gravitons, it follows that the the order parameter defines a rank 2 tensor $e_{ab}$ which describes the relative orientation of the paired gravitons. This suggests that the order parameter for the pairing wavefunction (1) can be identified with the metric for a 4-manifold. In fact this 4-dimensional manifold is uniquely determined as the 4-manifold whose null geodesics are defined by the intersection of the self-dual and anti-self-dual surfaces associated with the paired gravitons.

An elegant way to view this construction is to note that formally one can associate the self-dual and anti-self-dual pieces of the wavefunction (1) with curved twister spaces. The paired wavefunction then corresponds to a deformation of ambitwister space; i.e. elements (u,v) of $CP^3 \times CP^3$ such that u·v = 0. Thus the coherence of the wavefunction (1) ties together the contribution of each graviton pair to form a smooth manifold, and the Girvin-MacDonald order parameter for this system can be identified with the metric of a smooth 4-dimensional manifold. Furthermore, since the parts of the wavefunction corresponding to self-dual and anti-self-dual gravitons have opposite conformal weight we are led to the identification of the condensate wavefunction (1) with flat spacetime (the only asymptotically locally flat spacetime whose metric corresponds to a solution to eq. (2) with zero conformal weight is Minkowski space). An interesting aspect of this identification is that the pairing wavefunction (1) seems to correspond to flat rather than merely conformally flat spacetime, which may be a clue as to why the cosmological constant is zero.

It should be noted that the quantum system described by the wavefunction (1) resembles a superconductor in that there is a mass gap for excitations with respect to the time dimension of the underlying 2+1 dimensional theory. On the other hand there are Landau-Ginzberg-like collective excitations which correspond to slow variations in the order parameter. One can show [18] that slow variations in the order parameter satisfy the classical equations for dilatonic gravity. Therefore one expects that there will be massless s-wave excitations that can be identified with propagating modes of a Brans-Dicke like dilaton field, and massless d-wave excitations that can be identified with classical gravitational waves. The s-wave modes evidently correspond to variations in the density of graviton pairs, while the d-wave gravitational waves correspond to periodic variations in the relative orientation of the self-dual and anti-self-dual components of graviton pairs. This quantum model for spacetime is distinct from previous attempts to quantize gravity in that the underlying theory is formulated in 2+1 dimensions rather than 4 dimensions. The "stochastic" time $t_s$ of this underlying theory is not the same as the familiar time dimension of ordinary four dimensional spacetime. Indeed ordinary classical spacetime only emerges in the limit $t_s \to \infty$, corresponding to $H\Psi=0$. It might be noted in this regard that Gepner's construction of superstring vacua also suggests [13] that the vacuum states of superstrings can best be understood as the endpoint of relaxation with respect to an extra time dimension.

This quantum model of spacetime is a kind of generalization of string theory where self-dual and anti-self-dual Einstein spaces replace the left and right moving string modes, and a world volume replaces the world sheet. The self-dual and anti-self-dual Einstein spaces associated with the wavefunction (1) have a natural embedding in $C^3$ (or as we shall see in the next section $CP^3$). Therefore our quantum model of spacetime can also be described as a theory of 7-banes moving in a 13-dimensional spacetime. In the limit of infinite stochastic time the positions of these 7-banes become fixed, and thus in this limit the present theory might be considered to be a realization of Vafa's F-theory [3].

## 3. Icosahedral Quantum Gravity

Although the quantum model for spacetime described in section 2 does use the type of variables called for in a fundamental theory, it seems to describe dilatonic gravity but not matter. It also does not appear to exhibit the characteristic discrete symmetries associated with the Leech lattice. There is however a possible inconsistency in this model for spacetime in that if the Dirac strings that originate on the monopole singularities in the Gibbons-Hawking multi-instanton do not go off to infinity, but instead terminate on nearby anti-monopole singularities, then the global topology at infinity will no longer be nontrivial. This suggests that the model of ref. 4 is not completely consistent because the only self-dual or anti-self-dual solutions of the Einstein equations for which the global topology of spacetime at infinity is trivial is flat spacetime. The only obvious way to resolve this conflict is to assume that the self-dual and anti-self-dual pieces of the wavefunction (1) actually correspond to compact self-dual and anti-self-dual Einstein spaces. However, the only compact self-dual or anti-self-dual solutions of the vacuum Einstein equations are the 4-torus and the complex surface K3. The metric for the 4-torus contains no monopole-like singularities, and therefore is not of interest to us. The classical metric for a K3 manifold, on the other hand, contains 22 monopole-like singularities. Of these monopole-like singularities 16 come from the 8 Eguchi-Hanson instantons that are used [20] to replace the orbifold singularities in the construction of K3 from a 4-torus $Z_2$ orbifold. Thus it would probably be more consistent to regard the "gravitons" in our model of spacetime as pairs of K3 manifolds, and to suppose that classical spacetime arises from a coherent superposition of pairs of self-dual and anti-self-dual K3 manifolds.

More specifically we will introduce "icosohedral gravitons" consisting of a self-dual K3 manifold together with its mirror manifold paired with an anti-self-dual K3 manifold and its mirror. Together these four manifolds contain 32 pairs of monopole-like singularities. Obviously such a graviton has many internal degrees of freedom besides those needed for dilatonic gravity. We will now argue that these extra modes could very well correspond to the charged matter states of the heterotic string. Our argument is based on the natural assumption that the 32 pairs of monopole-like singularities in each graviton are arranged so that the 32 monopoles are spread over the surface of a 2-sphere in a more or less symmetrical fashion. In particular, we will suppose that the positions of the monopole singularities approximately correspond to the vertices and face centers of an icosahedron. Such an icosahedron will occur in each of the paired K3 manifolds that are used to construct our new model of the quantum gravity vacuum. The vertices and face centers of these two icosahedra will be connected by Dirac strings. In a manner analogous to our "ambitwister" interpretation of how curved four dimensional spacetime arises from the wavefunction (1) we will identify the excitations of the icosahedral vacuum state corresponding to four dimensional charged matter states with the distinct monodromies of monopole pairs inside an icosohedral graviton. Now the 64 monopole sigularities inside the paired K3 manifolds can be divided into 2016 pairs, with the position and orientation of each of these pairs being described by 6 parameters. In addition the fundamental theory has a global $Z_2$ symmetry corresponding to $\psi \to i\psi$, so that taking into account the four possible vacuum states for the paired K3 manifolds, the total number of diffrent types of low mass bosons would be 24 x 2016 = 12 x $2^{12}$ - 6 x 128. In the construction of the lowest dimensional non-trivial representation of the Monster group spaces of dimension 12 x $2^{12}$ play a crucial role, and so the occurence here of a space of excitations whose multiplicity differs from 12 x $2^{12}$ by a simple multiple of the number of massless states of a Type II superstring is circumstancial evidence that our icosahedral construction connects the Monster group with heterotic and Type II superstrings. Indeed it was conjectured in ref. 6 that the lowest nontrivial representation of the Monster group involved 12 heterotic and 6 Type II superstrings so that the 24 x 2016 internal excitations of our icosahedral graviton just matches the conjectured heterotic string contribution to the lowest nontrivial representation of the Monster. It should also be noted that our identification of heterotic-like states with deformations of paired icosahedra is quite analogous to the fact that string theories compactified on manifolds with nontrivial topologies can give rise to charged matter states when monopole-like singularities coalesce to form A-D-E type geometric singularities [21]. If the deformations depend on the coordinates of the lower dimensional space then the deformation parameters can be thought of as charged fields. Our derivation of heterotic-like matter states is similar, but is an improvement on these string model results in that the geometric pattern that gives rise to an enhanced gauge symmetry is not merely an ad hoc feature of the internal space, but is an intrinsic feature of the graviton itself.

If the coordinates of the paired K3 manifolds are identified, then the 2-dimensional null surface on which the fundamental fermions live can be thought of as an infinite momentum frame membrane embedded in an 11-dimensional spacetime (if we neglect the two coordinates associated with the $SU(\infty)$ degree of freedom). Thus the present theory might be regarded as a quantum version of M-theory. However as shown in ref. 12 the $SU(\infty)$ degree of freedom of the fundamental fermions can be treated as another dimension, so that the infinite momentum frame membrane is naturally associated with a 3-manifold. Therefore it is probably better to regard our theory as a quantum realization of F-theory. It is amusing that F-theory compactifications involving $K_3$ or equivalently orientifold type compactifications of Type IIB strings involving a 2-torus play a central role in our version of F-theory. As noted by Sen [22] an orientifold compactification of a Type II string on a 2-torus is dual to an ordinary compactification of a heterotic string on the 2-torus, thus supporting our surmise that string-string duality in eight dimensions is an important clue as to the nature of the underlying theory.

It should be kept in mind, though, that the self-dual and anti-self-dual 4-manifolds in our theory and their associated null surfaces and 3-geometries evolve with respect to stochastic time rather than ordinary time. Furthermore our theory describes an infinite ensemble of 4-manifolds rather than a single membrane or 3-bane. It is interesting to note, though, that for a fixed value of stochastic time the statistical partition function for an infinite assembly of membranes is equivalent to a Polyakov-like path integral for a single string propagating in a 10-dimensional space. Likewise for fixed stochastic time the statistical partition function for an infinite sum of Euclidean 4-manfiolds becomes a Hawking-like path integral for a single 4-manifold. Thus Euclidean quantum cosmology will emerge from our theory of interacting $K_3$ manifolds in much the same way that ordinary Euclidean quantum mechanics emerges from a theory of interacting strings [23].

## 4. Conclusion

We have noted that the S, T, and U duality symmetries and string-string duality of superstring models in six and eight dimensions are isomorphic to elementary group structures that play an important role in the construction of the Monster sporadic group. In particular string-string duality in these dimensions seems to be identical to an involution that has played an important role in constructions of the Monster group. The explicit appearence of this involution symmetry in six and eight dimensions suggests that a twelve dimensional formulation of superstring theory should exist in which the fundamental quantum degrees of freedom can be identified with elements of a commutative non-associative algebra. Thus we were led to a formulation of the 12-dimensional theory as the ground state of a theory of fermions moving in two dimensions and interacting via an $SU(\infty)$ gauge potential.

The 12-dimensions arise because a 2+1 dimensional theory of fermions interacting via an $SU(\infty)$ gauge potential has a natural interpretation as an ambitwister theory of a curved four dimensional spacetime. In particular the 2-dimensional surfaces on which the fermions move correspond to a null surfaces in four dimensional self-dual or anti-self-dual Einstein spaces. The space of all such surfaces is a deformation of $CP^3 \times CP^3$, and real four dimensional spacetime is interpreted as the diagonal space of null geodesics formed by the intersections of the null surfaces. In ref. 4 it was noted that this formulation for quantum gravity contains gravitons and dilatons, and in this paper it was shown that if the Einstein spaces are taken to be K3 manifolds then heterotic-like matter degrees of freedom are also included. In a sense one can regard the emergence of matter-like degrees of freedom in our theory as an elaborate example of the phenomenom of enhanced gauge symmetry that occurs in compactified superstring theories when the singularities of an internal space coelecse to form a symmetrical geometrical pattern.

The possibility of relating a fundamental theory of superstrings to properties of the Leech lattice and the Monster group is certainly intriguing from the point of view that in many respects the Leech lattice and its associated symmetry groups are mathematically unique. This uniqueness can be traced to the fact first noticed by Galois that only in certain exceptional cases does the group $L_2(q)$ of linear fractional transformations over the finite field $F_q$ have a representation with dimension smaller than q+1. These exceptional representations play a fundamental role in the existence of the Leech lattice and its associated symmetry groups. The fact that these same exceptional representations occur as multiplets of superstring BPS states in certain dimensions is therefore strong circumstantial evidence that the unique properties of sporadic finite groups play a significant role in the theory underlying superstring models. The occurence of coincidences between discrete duality symmetries of superstring states and linear fractional groups with exceptional representations also supports the conjecture made in reference 6 that the massless states of the heterotc and Type II superstrings can be identified with the words of the Golay code. In fact the existence of the exceptional representations of $L_2(q)$ is closely related to the error correcting property of the Golay code, and the exceptional representations of $L_2(q)$ are clearly visible as subgroups of the permutation group for the Golay code.

As a final remark we would like to point out the theory we have described provides a connection between the topology of self-dual 4-manifolds, superstring models, and the Monster sporadic group. The intersection matrix for K3 manifolds can be represented in the form:

$$-E_8 \oplus -E_8 \oplus 3\begin{pmatrix} 0 & 1 \\ 1 & 0 \end{pmatrix}.$$

Evidently the piece of this intersection matrix involving two $E_8$'s plays much the same role in our theory as the gauge group of the $E_8 \otimes E_8$ heterotic string. Indeed we showed in section 3 that the monopole singularities associated with these 2-cycles can be used to construct the piece of the smallest non-trivial representation of the Monster group that appears to be associated with heterotic superstrings. The possibility of a connection between the Monster group and the topology of K3 manifolds has not been previously suspected, but such a connection is undoubtedly related to "Monsterous Moonshine." One possibility for explaining this connection is to note that in the limit $t_s \rightarrow \infty$ the underlying equations of our theory describe a gauge fixed version of a 2-dimensional topological gauge theory which is known [24] to be related to both the properties of Riemann surfaces and the topological properties of 4-manifolds. From this point of view our theory appears to be closely related to the demonstration by Sieberg and Witten that supersymmetric Yang-Mills-Higgs theory can be used to understand the topology of 4-manifolds [25].

The author is grateful for the hospitality of the Stanford University Physics Department during the 1995-96 academic year.

# References

1. J.H. Schwarz, "An SL(2,Z) Multiplet of Type IIB Superstrings", Phys. Lett. 360B (1995) 13; "The Power of M Theory" hep-th/9510086.
2. P. Horava and E. Witten, "Heterotic and Type I String Dynamics From Eleven Dimensions", Nucl. Phys. B460 (1996) 506.
3. C. Vafa, "Evidence for F Theory" hep-th/9602022.
4. G. Chapline, "Quantum Model for Spacetime", Mod. Phys. Lett. A7, (1992) 1959; "Anyons and Coherent States for Gravitons",in <u>Proceedings of the XXI International Conference on Differential Geometric Methods in Theoretical Physics,</u> edited by C. N. Yang, M. L. Ge, and X. W. Zhou (World Scientific, Singapore,1993).
5. R. L. Griess, "The Friendly Giant", Invent. Math. 69 (1982) 1.
6. G. Chapline, "Unification of Gravity and Elementary Particle Interactions in 26 Dimensions", Phys. Lett. 158B (1985).
7. C. M. Hull and P.K. Townsend, "Unity of Superstring Dualities", Nucl. Phys. B438 (1995) 109.
8. J.H. Conway, "A simple construction for the Fischer-Griess monster group", Invent. Math. 79 (1985) 513.
9. I.B. Frenkel, J. Lepowsky, and A. Meurman, "A natural representation of the Fisher-Griess Monster with the modular function J as character", Proc. Natl. Acad. Sci. <u>81</u> (1984) 3256.
10. E. Witten, "Dynamics of String Theory in Various Dimensions", Nucl. Phys. B443 (1995) 573.
11. S. M. Girvin, et. al., "Exactly Soluble Model of Fractional Statistics", Phys. Rev. Lett. 65, 1671 (1990).
12. G. Chapline and K. Yamagishi, "Three Dimensional Generalization of Anyon Superconductivity", Phys. Rev. Lett. 66, 3064 (1991).
13. G. Chapline, "Superstrings and Quantum Gravity", Mod. Phys. Lett. A5 (1990) 2165.
14. C. Vafa. and N. Warner, "Catastrophes and the Classification of Conformal Theories", Phys. Lett. 218B (1989) 51.
15. S. M. Girvin and A.H. MacDonald, "Off-Diagonal Long Range Order, Oblique Confinement, and the Fractional Quantum Hall Effect", Phys. Rev. Lett. 58 (1987) 1252 .
16. N. Seiberg and E. Witten, "Electric-Magnetic Duality, Monopole Condensation, and Confinement in N=2 Supersymmetric Yang-Mills Theory",  Nucl. Phys. B426 (1994) 19.
17. G. W. Gibbons and S. W. Hawking, "Gravitational Multi-Instantons", Phys. Lett. 78B, 430 (1978).

18. This can be seen by examining the double commutator for covariant derivatives $[ D_x{}^a + D_y{}^a , [ D_x{}^a + D_y{}^a , D_x{}^b + D_y{}^b ] ]$, where $x^a$ and $y^a$ are coordinates on the 4-dimensional self-dual and anti-self-dual Einstein spaces corresponding to the self-dual and anti-self-dual gravitons; cf. C. LeBrun, Comm. Math. Phys. 139, 1 (1991).
20. D.N. Page, "A Physical Picture of the K3 Gravitational Instanton", Phys. Lett. 80B (1978) 55.
21. M. Bershadsky, et. al., "Geometric Singularities and Enhanced Gauge Symmetries", hep-th/9605200.
22. A. Sen, "F-theory and Orientifolds", hep-th/9605150.
23. S-Y. Chu, "Statistical Origin of Classical Mechanics and Quantum Mechanics", Phys. Rev. Lett. 71 (1993) 2847.
24. G. Chapline and B. Grossman, "Conformal Field Theory and a Topological Quantum Theory of Vortices", Phys. Lett. 223B (1989) 336.
25. E. Witten, "Monopoles and 4-manifolds", Math. Res. Letters 1 (1994) 764.